\documentclass{PoS}

\usepackage{txfonts}
\usepackage{mathrsfs}

\newcommand{\bfl}{\begin{flushleft}}
\newcommand{\efl}{\end{flushleft}}

\newcommand{\ud}{\mathrm{d}} 
\newcommand{\be}{\begin{equation}}
\newcommand{\ee}{\end{equation}}

\newcommand{\zhh}{$\mathrm{\zeta^{H_{2}}}$}

\newcommand{\Ezhh}{\mathrm{\zeta^{H_{2}}}}

\newcommand{\ea}{$\mathrm{\eta_{AD}}$}
\newcommand{\eh}{$\mathrm{\eta_{H}}$}
\newcommand{\eo}{$\mathrm{\eta_{O}}$}

\title{Cosmic-ray propagation at small scale: a support for protostellar disc formation}

\ShortTitle{Cosmic rays and protostellar disc formation}

\author{\speaker{Marco Padovani}$^{abc}$
, D. Galli$^{c}$, P. Hennebelle$^{d}$, B. Commer\c con$^{e}$, and M. Joos$^{d}$\\
\llap{$^{a}$} Laboratoire Univers et Particules de Montpellier, UMR 5299 du CNRS, Universit\'e de Montpellier, place E. Bataillon, cc072, 34095 Montpellier, France\\
\llap{$^{b}$} Laboratoire de Radioastronomie Millim\'etrique, UMR 8112 du CNRS, \'Ecole Normale Sup\'erieure et Observatoire de Paris, 24 rue Lhomond, 75231 Paris cedex 05, France\\
\llap{$^{c}$} INAF--Osservatorio Astrofisico di Arcetri, Largo E. Fermi 5, 50125, Firenze, Italy\\
\llap{$^{d}$} CEA, IRFU, SAp, Centre de Saclay, 91191 Gif-Sur-Yvette, France\\
\llap{$^{e}$} \'Ecole Normale Sup\'erieure de Lyon, CRAL, UMR 5574 du CNRS, Universit\'e Lyon I, 46 All\'ee d'Italie, 69364 Lyon cedex 07, France\\
E-mail: \email{Marco.Padovani@umontpellier.fr}, \email{galli@arcetri.astro.it},
\email{hennebelle@cea.fr}, \email{benoit.commercon@ens-lyon.fr}, \email{joos@cea.fr}
}

\abstract{As long as magnetic fields remain frozen into the gas, the magnetic braking prevents the formation of protostellar discs.
This condition is subordinate to the ionisation fraction characterising the inmost parts of a collapsing cloud.
The ionisation level is established by the number and the energy of the cosmic rays able to reach these regions. 
Adopting the method developed in our previous studies, we computed how cosmic rays are attenuated as a function of column density and magnetic field strength. We applied our formalism to low- and high-mass star formation models obtained by numerical simulations of gravitational collapse that include rotation and turbulence.
In general, we found that the decoupling between gas and magnetic fields, condition allowing the collapse to go ahead, occurs only when the cosmic-ray attenuation is taken into account
with respect to a calculation in which 
the cosmic-ray ionisation rate is kept constant.
We also found that the extent of the decoupling zone also depends on the dust grain size distribution and is larger 
if large grains (of radius $\sim 10^{-5}$~cm) are formed by compression
and coagulation during cloud collapse. The decoupling region disappears for the high-mass
case due to magnetic field diffusion that is caused by turbulence
and that is not included in the low-mass models.
We infer that a simultaneous study of the cosmic-ray propagation during the cloud's collapse
may lead to values of the gas resistivity in the innermost few hundred AU around a forming protostar
that is higher than generally assumed.
}

\FullConference{Cosmic Rays and the InterStellar Medium\\
		 24-27 June 2014\\
		 Montpellier, France}

\begin{document}

\section{Introduction}
\label{intro}

Cosmic rays (hereafter CRs) have a twofold role in the interstellar medium. In fact, they both originate the chemistry in molecular clouds and
have a fundamental function in controlling the collapse of a cloud and the resulting star formation.
In prestellar and Class~0 sources, CRs represent the main ionising agent, since X-rays ionisation arises only in presence of 
embedded young stellar objects~\cite{kk83,sn83} and interstellar UV photons are absorbed for a visual extinction $A_V\gtrsim 4$~magnitudes~\cite{m89}.
During the collapse of a cloud, magnetic fields coupled to the gas cause a brake of any rotational 
motions, at least as long as the field remains frozen
into the gas and the rotation axis of the cloud is close to the mean
direction of the field, e.g.~\cite{gl06,ml08,hf08}. This is why the study of the formation
of circumstellar discs still presents theoretical challenges.
However, discs
around Classes I and II young stellar objects are commonly observed~\cite{wc11,ts12},
and there is also some evidence of discs around Class 0 objects~\cite{th12,mu13}.

In order to alleviate the magnetic braking, different mechanisms have been introduced:
({\em i}\/) non-ideal magnetohydrodynamic (MHD) effects~\cite{sg06,db10,kl11,bw12a,bw12b};
({\em ii}\/) misalignment between the main magnetic field direction and the rotation axis~\cite{hc09,jh12}; 
({\em iii}\/) turbulent diffusion of the magnetic field~\cite{sb12,sd13,jh13};
({\em iv}\/) flux redistribution driven by the interchange instability~\cite{kl12};
and ({\em v}\/) depletion of the infalling envelope anchoring the magnetic field~\cite{ml09,mi11}.

Non-ideal MHD effects, namely ambipolar, Hall, and Ohmic diffusion,
depend on the abundances of charged species as well as on their mass and
charge. The ionisation fraction, in turn, is determined by CRs
in cloud regions of relatively high column density 
where star formation takes place.
The CR ionisation rate (\zhh) is usually assumed to be equal to a
``standard'' (constant) value of $\zeta^{{\rm H}_2}\approx 10^{-17}$~s$^{-1}$~\cite{st68}. 
However, CRs interacting with $\mathrm{H_{2}}$ 
in a molecular cloud lose energy by several processes, mainly
by ionisation losses (see~\cite{pgg09}, hereafter PGG09).
As a consequence, while low-energy CRs ($E\lesssim100$~MeV)
are possibly prevented from entering a molecular cloud because of streaming instability~\cite{cv78,mg14}, high-energy CRs are slowed down
to energies that are relevant for ionisation (ionisation cross sections for protons and electrons colliding with
$\mathrm{H_{2}}$ peak at about 10~keV and 0.1~keV, respectively).  

It has been shown in PGG09 that \zhh\ can decrease by about two orders of magnitude from diffuse regions of column densities
$\sim 10^{21}$~cm$^{-2}$ ($\Ezhh\sim10^{-16}-10^{-15}$~s$^{-1}$) to dense cores and massive protostellar envelopes
with column densities of $\sim 10^{24}$~cm$^{-2}$ ($\Ezhh\sim10^{-18}-10^{-17}$~s$^{-1}$).
A further attenuation is caused by the presence of magnetic fields. In fact, a poloidal field threading a molecular cloud core
reduces \zhh\ on average by a factor 3--4 depending on the magnetisation degree and the position inside the core~\cite{pg11,pg13}.
A stronger reduction of \zhh\ takes place in the inner region of a core, where the formation of a protostellar disc is expected
to occur ($\lesssim100-200$~AU) because the toroidal field component generated by rotation boosts the path length of particles~\cite{phg13}.

In order to properly treat the problem of the influence of CRs on the collapse dynamics and the circumstellar disc formation, one should 
compute the CR propagation self-consistently with the evolution of magnetic field and density, following at the same time the formation
and the destruction of chemical species.
Since this approach would be extremely time-consuming from a numerical point of view, 
$(i)$ we take some snapshots of density and magnetic field
configuration from ideal MHD simulations that do not include any gas resistivity; 
$(ii)$ we follow the propagation of CRs, computing their spatial distribution and the CR ionisation rate;
$(iii)$ using a simplified chemical model, we approximately calculate the chemical composition as a function of the values of 
\zhh\ computed in the previous step; $(iv)$ finally, we compute the microscopic resistivities (ambipolar, Hall, and Ohmic)
and compare the 
time scale of magnetic field diffusion $t_B$ to the dynamical time scale $t_{\rm dyn}$ 
at each point in the model to determine the region of magnetic decoupling, where $t_B <t_{\rm dyn}$.
The hypothesis of ideal MHD on which the simulations are based becomes invalid by definition inside the decoupling region.
However, even if our calculations are not fully self-consistent, we show that assuming the proper value of \zhh\
at different depths, gives strong constraints on the disc formation.

\section{Energy loss processes and magnetic effects on cosmic-ray propagation}
While passing through a molecular cloud, a CR undergoes collisions with molecular hydrogen. According to its energy and
composition, it is slowed down 
due to processes that are specific of a particular kind of particle (bremsstrahlung, synchrotron emission, and inverse Compton scattering for electrons; elastic interactions, pion production, and spallation for protons) or common both to CR protons and electrons (Coulomb and inelastic interactions, and ionisation). 
PGG09 show that even if a local CR interstellar spectrum is lacking of low-energy particles, 
the slowing-down of high-energy CR protons and electrons during their propagation produces a low-energy tail. 
Our modelling is able to explain the decrease of \zhh\ with increasing hydrogen column density computed from observations. In particular, a proton component at low energies, and most likely also an electron component, could be necessary to reproduce the data.

The quantity that describes the energy losses during the propagation is called {\em energy loss function} and it is given by
\be\label{elossf}
L_{k}(E_{k}) = -\frac{\ud E_{k}}{\ud N({\rm H_{2}})}\,,
\ee
where $N{\rm(H_{2})}$ is the column density of the medium in which the particle of species $k$ and energy $E_{k}$ propagates
(see e.g. left panel of Fig.~3 in~\cite{pg13}).
Besides energy losses, one has to account for the fact that CRs are charged particles and they move along the field lines
following an helicoidal path. This means that they ``see'' a larger H$_{2}$ column density with respect to a rectilinear propagation, so that
the column density in Eq.~(\ref{elossf}) reads
\be
N(\alpha) = \int_{0}^{\ell_{\rm max}(\alpha)}n(\ell)\ud\ell\,,
\ee
where $\ell_{\rm max}$ is the maximum depth reached inside the core and $n(\ell)$ is the H$_{2}$ volume density.
The angle $\alpha$, called {\em pitch angle}, is the angle between the CR velocity and the direction of the magnetic field. 
Its evolution during the CR propagation reads
\be
\alpha=\arccos\sqrt{1-\chi+\chi\cos^{2}\alpha_{\rm ICM}}\,,
\ee
where $\chi$ is the ratio between the local and the interstellar magnetic field and $\alpha_{\rm ICM}$ is the initial pitch angle 
(see~\cite{pg11} for more details).
Magnetic focusing and magnetic mirroring are the two competing effects arising in presence of a magnetic field. 
The former increases the CR flux where the field is more concentrated, enhancing \zhh, the latter bounces CRs out
of the cloud as long as $\alpha\rightarrow\pi/2$, namely when the CR velocity is perpendicular to the magnetic field line.
A simple fitting formula which combines in a single expression the effects of energy losses and magnetic fields for different CR
proton and electron interstellar spectra is given by~\cite{phg13}:
\be
\label{fitfor}
\zeta_{k}^{\rm H_{2}}(\alpha)=\frac{\zeta_{0,k}^{(\rm low~N)}\zeta_{0,k}^{(\rm high~N)}}%
{\zeta_{0,k}^{(\rm high~N)}\left[\displaystyle{\frac{N(\alpha)}{10^{20}~{\rm cm}^{-2}}}\right]^{a}+%
\zeta_{0,k}^{(\rm low~N)}\left[\exp\left(\displaystyle{\frac{\Sigma(\alpha)}{\Sigma_{0,k}}}\right)-1\right]}\,,
\ee 
where $\Sigma(\alpha)=\mu m_{\mathrm p}N(\alpha)/\cos\alpha$
is the effective surface density seen by a CR propagating with pitch
angle $\alpha$, $m_{\mathrm p}$ the proton mass and $\mu = 2.36$
the molecular weight for the assumed fractional abundances of H$_{2}$
and He. The fitting coefficients of Eq.~(\ref{fitfor}) are given
in~\cite{pgg09,pgg13}.

\section{Cosmic-ray ionisation rate at high densities}
At higher column densities ($N\gtrsim10^{25}$~cm$^{-2}$) the drop in \zhh\ becomes even more dramatic, since the CR attenuation starts to be
exponential. Besides, the stronger the toroidal field component, the larger will be the path travelled by a particle.
A useful fitting formula to evaluate the effective column density covered by a charged particle is given by~\cite{phg13}.
If $N(\mathrm{H_{2}})$ is the average column density seen by an
isotropic flux of CRs, $N_{\rm eff}$ has the form
\be
\label{Neff}
N_{\rm eff} = (1 + 2\pi\ {\cal F}^{s})\ N(\mathrm{H_{2}})\,.
\ee
The factor ${\cal F}$ depends on the ratio between the toroidal and the poloidal components of
the magnetic field, $b=|B_{\varphi}/B_{p}|$, as well as on its module. It reads
\be\label{calP}
{\cal F}=\frac{|{\bf B}|}{1\ \mu{\rm G}}\frac{\sqrt{b^{*}}}{2}\,,
\ee
where
\be\label{betatilde}
b^{*}=\frac{b-b_{\rm min}}{b_{\rm max}-b_{\rm min}}\,,
\ee
$b_{\rm min}$ and $b_{\rm max}$ being the minimum and the maximum
value of $b$ in the whole data cube, respectively.  When the magnetic
field strength is negligible, CRs propagate along straight lines.
In this case ${\cal F}=0$ and $N_{\rm eff}=N({\rm H_{2}})$, otherwise
${\cal F}>0$ and $N_{\rm eff}>N({\rm H_{2}})$. 
Besides, the higher the density, the stronger is the role of the magnetic
field in increasing $N_{\rm eff}$. This justifies the presence in
Eq.~(\ref{Neff}) of the power $s$ that reads
\be
\label{spower}
s \simeq 0.7\frac{\log_{10}(n/n_{\rm min})}{\log_{10}(n_{\rm max}/n_{\rm min})}\,,
\ee
$n_{\rm min}$ and $n_{\rm max}$ being the minimum and the maximum
value of the density in the whole data cube, respectively.
Notice also that the factor ${\cal F}$ depends both on the local value of the
magnetic field (through $b$ and $|{\bf B}|$) and on the large scale
configuration (by means of $b^{*}$ and $s$).
Once evaluated the effective column density, the corresponding
(effective) CR ionisation rate, $\zeta^{\rm H_{2}}_{\rm eff}$, is
obtained by
\be
\label{zetaeff}
\zeta^{\rm H_{2}}_{\rm eff}=\kappa \zeta^{\rm H_{2}}(N_{\rm eff})\,,
\ee
where $\zeta^{\rm H_{2}}(N_{\rm eff})$ is computed using
Eq.~(\ref{fitfor}) after replacing $N(\alpha)$ and $\Sigma(\alpha)$
with $N_{\rm eff}$ and $\mu m_{\rm H}N_{\rm eff}$, respectively.
The factor $\kappa$ is given by
\be\label{kappa}
\kappa = \frac{1}{2}+\frac{1}{\pi}\arctan\left(\frac{800\ \mu{\rm G}}{|{\bf B}|}\right)
\ee
and it represents the correction for magnetic effects.

\section{Diffusion coefficients and time scales}
Ions and electrons are frozen into magnetic field, while neutrals can diffuse through reaching the central part of the core. 
A frictional force couples charged particles and neutrals. Since CRs regulate the ionisation degree, they set limits on the coupling between gas and magnetic field, controlling the time scale of the collapse.
In general, at lower densities ($n_\mathrm{H_{2}}\lesssim10^{8}-10^{9}$~cm$^{-3}$) the diffusion is controlled by the ambipolar resistivity.
At intermediate densities ($10^{8}-10^{9}$~cm$^{-3}\lesssim
n_\mathrm{H_{2}}\lesssim10^{11}$~cm$^{-3}$) Hall diffusion dominates: the more massive charged species (molecular ions and charged grains)
are coupled with the neutral gas through collisions and then they are decoupled from magnetic field.
At the highest densities
($n_\mathrm{H_{2}}\gtrsim10^{11}$~cm$^{-3}$,~\cite{un81}) ions and electrons are knocked off from field lines due to collisions and the 
Ohmic dissipation sets in.
However, the extent to which a diffusion process dominates over the
others hinges on several factors, one of which is the assumed grain size distribution (see e.g. Fig.~4 in~\cite{pgh14}) as
well as the CR ionisation rate. For this reason,
after determining the variation of \zhh\ as a function of the density distribution and magnetic field configuration,
we investigated about how microscopic resistivities are affected by remarkable deviation 
of the CR ionisation rate from a constant value.

The induction equation reads
\be
\label{dBdt} 
\frac{\partial\vec B}{\partial t}+\nabla\times(\vec B\times\vec U) 
= \nabla\times\left\{\eta_\mathrm{O}\nabla\times\vec
B+\eta_\mathrm{H}(\nabla\times\vec B)\times\frac{\vec B}{B}
+\eta_\mathrm{AD}\left[(\nabla\times\vec B)\times\frac{\vec B}{B}\right]\times
\frac{\vec B}{B}\right\}\,, 
\ee
where $\vec U$ is the fluid velocity and $\vec B$ the magnetic field
vector. Ambipolar, Hall, and Ohmic resistivities (\ea, \eh, and
\eo, respectively) can be written as a function of the parallel
($\sigma_{\parallel}$), Pedersen ($\sigma_{\rm P}$) and Hall
($\sigma_{\rm H}$) conductivities (e.g.~\cite{w07,pgb08,pgh14}). These conductivities depend on 
the ionisation fraction that, in turn, is determined by \zhh.
We calculated the ionisation fraction of all the species involved (electrons, metal and molecular ions, neutral and charged grains,
H$^{+}$ and H$_{3}^{+}$) using the chemical model described in Appendix~A in~\cite{pgh14}.

The drift velocity of the magnetic field ($\vec U_{B}$) can be represented by the velocity of the charged species, which are frozen with field lines, with respect to neutrals. From the comparison of this velocity with the fluid velocity, 
it is possible to assess the degree of diffusion of the field and then to estimate the size of the region where gas and magnetic field are decoupled.
$\vec U_{B}$ can be written as
a function of resistivities~\cite{nn02}, 
allowing to isolate the ambipolar (AD), Hall (H), and Ohmic (O)
contributions, namely
\be 
\vec U_{B}=\vec U_{\rm AD}+\vec U_{\rm H}+\vec U_{\rm O}\,, 
\ee
where 
\be 
\vec U_{\rm AD} = \frac{4\pi\ \eta_{\rm
AD}}{cB^{2}}\vec j\times\vec B\,,\quad \vec U_{\rm H} = \frac{4\pi\
\eta_{\rm H}}{cB^{3}}(\vec j\times\vec B)\times\vec B\,,\quad \vec
U_{\rm O} = \frac{4\pi\ \eta_{\rm O}}{cB^{2}}\vec j\times\vec B\,.
\ee 
Then 
\be 
\vec U_B=\frac{4\pi}{cB^{2}}\left[\left(\eta_{\rm AD}+\eta_{\rm
O}\right)\vec j\times\vec B+\eta_{\rm H}\left(\vec j\times\vec
B\right)\times\frac{\vec B}{B}\right]\,,  
\ee 
where $\vec j=(c/4\pi)\nabla\times\vec B$ is the current density.
Thus, the diffusion time of the magnetic field,
$t_B$, can be written as a function of the time scales associated to the three diffusion
processes, 
\be 
\frac{1}{t_B}=\frac{1}{t_{\rm AD}}+\frac{1}{t_{\rm H}}+\frac{1}{t_{\rm O}}\,,
\ee 
where $t_k=R/U_k$ ($k=\mathrm{AD,H,O}$) and $R$ a typical length scale of the region.  
The diffusion time of the magnetic field can 
then be compared to the time scale of evolution of the fluid, $t_{\rm dyn}$. 
We define the dynamical time scale of the cloud as $t_{\rm dyn}=R/U$, where $U$
is the fluid velocity, including both infall and rotation. 
In regions where $t_{B}<t_{\rm dyn}$  
the magnetic field is partially decoupled and therefore has less influence on the gas dynamics while, if $t_{B}>t_{\rm
dyn}$, diffusion is not efficient enough and the magnetic field remains well
coupled to the gas.

We performed our calculations for three snapshots from numerical simulations: two low- and one high-mass cases.
We adopted both a 
spatially uniform CR ionisation rate of $\Ezhh=5\times10^{-17}$~s$^{-1}$,
and the formalism described in the previous sections 
to evaluate the attenuation of 
CRs in a magnetised cloud. We found that the ionisation fraction is 
significantly lower in the second case, and therefore the coupling with 
the magnetic field is weaker than usually assumed in the central region
of a collapsing cloud (see Fig.~5--7 in~\cite{pgh14}).

\section{Conclusions}
We applied our modelling for cosmic-ray propagation from large (0.1~pc) to small ($<100$~AU) scales 
where the formation of a protostellar disc is expected. We demonstrated that CRs play a lead role in regulating
the decoupling between the gas and the magnetic field and then in the cloud collapse.
As an instance, a decoupling zone of radius of $50-100$~AU around the central protostar is found in the case
with variable \zhh, but not when \zhh\ is assumed constant for a low-mass case snapshot. Besides, this size compares well with the
size of the protostellar discs (see Fig.~6 in~\cite{pgh14}).
Also in the high-mass case we examined (see Fig.~7 in~\cite{pgh14}) a decoupling zone of size $\sim 100$~AU is found.
However, its size becomes smaller for smaller grains, and 
disappears altogether for grains smaller than $10^{-6}$~cm.

Although the models adopted do not represent a time sequence, they nevertheless 
suggest that a decrease in the ionisation and/or an increase in the resistivity occurs 
in the innermost region of a cloud some time 
after the onset of collapse, but not earlier. In fact, the conditions for a substantial increase in the magnetic 
diffusion time are that the field is considerably twisted
and that dust grains had time to grow by coagulation. 
It is tempting to speculate that large, 100~AU-size discs are only allowed to form at a later stage when
the powerful magnetic brake on the infalling gas has been relieved by either (or a combination) of these effects.

Finally, in addition to \zhh, 
the size of the decoupling zone is also determined by the grain size assumed in the chemical model
and the volume of this region decreases for smaller grain size. The maximum grain size assumed in our work
($a_{\rm min}=10^{-5}$~cm) is likely a realistic value for the condition
expected in disc-forming regions, where larger grains are predicted
to form by compression and coagulation of smaller grains. 

\acknowledgments
MP and PH acknowledge the financial support of the Agence National
pour la Recherche (ANR) through the COSMIS project. 
MP and DG also acknowledge the support of the CNRS-INAF PICS project ``Pulsar wind nebulae,
supernova remnants and the origin of cosmic rays''.
This work has been carried out thanks to the support of the OCEVU Labex (ANR-11-LABX-0060) 
and the A*MIDEX project (ANR-11-IDEX-0001-02) funded by the ``Investissements d'Avenir'' French government programme managed by the ANR.


\end{document}